\documentclass[conference,compsoc]{IEEEtran}

\usepackage{color}
\usepackage{verbatim}
\usepackage{afterpage}
\usepackage{url}
\usepackage{xfrac}
\usepackage{algorithm}
\usepackage{algorithmicx}
\usepackage{algpseudocode}
\usepackage{amsmath}
\usepackage{amsfonts}
\usepackage{amssymb}
\usepackage{braket}
\usepackage{graphicx}
\usepackage{braket}
\usepackage[nocompress]{cite}
\usepackage{booktabs}
\usepackage{multirow}
\usepackage{caption}
\usepackage{subcaption}
\usepackage{cite}

\long\def\comment#1{}

\def\parah#1{\vspace*{0.0in} \noindent{\bf #1:}}

\newcommand{\I}{\mathrm{i}\mkern1mu}
\newcommand{\cn}{\textsc{CNOT\ }}
\IEEEoverridecommandlockouts

\begin{document}

\title{QFAST: Conflating Search and Numerical Optimization for
  Scalable Quantum Circuit Synthesis}

\author{
	\IEEEauthorblockN{Ed Younis\IEEEauthorrefmark{1},
					  Koushik Sen\IEEEauthorrefmark{2},
					  Katherine Yelick\IEEEauthorrefmark{2},
					  Costin Iancu\IEEEauthorrefmark{1}}
	\IEEEauthorblockA{\IEEEauthorrefmark{1}Computational Research Division\\
										   Lawrence Berkeley National Laboratory\\
					  \IEEEauthorrefmark{2}Department of Electrical Engineering 
										   and Computer Science\\
										   University of California Berkeley\\
										   Berkeley, CA}
}

\maketitle

\begin{abstract}
  We present a quantum synthesis algorithm designed to produce short
  circuits and to scale well in practice. The main contribution is a
  novel representation of circuits able to encode placement and
  topology using generic ``gates'', which allows the QFAST algorithm
  to replace expensive searches over circuit structures with few 
  steps of numerical optimization.  When compared against optimal
  depth, search based state-of-the-art techniques, QFAST produces
  comparable results:  $1.19\times$ longer  circuits up to four
  qubits, with an increase in compilation speed of $3.6\times$.  In
  addition, QFAST scales up to seven qubits.   When compared with the state-of-the-art ``rule'' based
  decomposition techniques in Qiskit, QFAST produces circuits shorter
  by up to  two orders of magnitude ($331\times$), albeit $5.6\times$ slower.
  We also demonstrate the composability with other techniques and the tunability of our formulation in
  terms of circuit depth and running time. 
\end{abstract}


\section{Introduction}

Quantum synthesis techniques generate circuits from high level
mathematical descriptions of an algorithm. Thus, they can provide a
very powerful tool for circuit optimization, hardware design
exploration and algorithm discovery. An important quality metric of
synthesis and compilers in general is circuit depth, as it relates
directly to the program performance. Short depth circuits are
especially important for Noisy Intermediate Scale Quantum (NISQ) era
devices, characterized by limited coherence time and noisy gates: here
synthesis can morph from a capacity into a capability provider. 

Synthesis has a long and distinguished research
history~\cite{ola15,ZXZ16,DawsonNielson05,MIM13,ctmq,kmm13,ionsynth,tucci2005kak,shende2006synthesis,isometries2016iten,qsearch}. At one end of the spectrum, optimal depth
algorhithms have been introduced for two qubits~\cite{tucci2005kak} (KAK),
augmented very recently with search based techniques~\cite{qsearch}
that ``scale''\footnote{Compilation finishes in a ``reasonable''
  amount of time.} up to four qubits. The latter employ a bottom-up
approach, where the solution circuit is extended one layer at a time
while taking into account the physical topology of the target Quantum
Processing Unit (QPU).
At the other end of the spectrum, top-down approaches~\cite{shende2006synthesis,isometries2016iten} use a ``rule''
based, divide-and-conquer decomposition approach. An exponent
technique is deployed in
IBM Qiskit~\cite{qiskit}, which uses linear algebra inspired, top-down decomposition. While more scalable, top-down
techniques lack a good solution for topology awareness and tend to
generate much longer (orders of magnitude) circuits than the slower
optimal techniques.  

The behavioral quality of optimal techniques is determined by several
factors. The search algorithm determines the number of partial
solutions evaluated: any strategy to reduce these will improve
scalability.
Each partial solution is subject to numerical optimization, which in turn
scales exponentially with the number of parameters and it is greatly
affected by the formulation of the objective function. Thus, these bottom-up
techniques slow down the closer they get to a solution. Furthermore,
topology awareness greatly reduces the final solution depth.

Quantum Fast Approximate Synthesis Tool (QFAST) has been designed to
improve scalability by trading-off search computational complexity
with numerical optimization complexity. It implements a bottom-up
approach, whose main contribution comes from encoding and using ``powerful'' computational building
blocks. Most search based techniques~\cite{qsearch} use
simple 2-qubit  parameterized blocks that contain two $U3$ gates and a
two-qubit gate, usually $CNOT$\footnote{For superconducting
  qubits.}.

In contrast, QFAST can use arbitrary generic gates applied
to any arbitrary set of qubits. The generic gates in QFAST encode 
{\it function}, i.e. the quantum transformation performed by the gate,
together with {\it location}, i.e. the set of qubits operated on.
QFAST  uses a three stage topology aware hierarchical algorithm. First, the
circuit is built bottom-up by adding one generic gate at a time. For a
$n$-qubit input algorithm, we start by adding  $m$-qubit generic gates, $m <
n$. Initially, this
gate encodes the application of any $m$-qubit program on any subset of
$m$ qubits from the orignal $n$-qubit circuit. A numerical optimization subroutine specializes the gate to a certain computation
on a single set of $m$-qubits such as we make most progress to
solution (minimize a unitary distance function). The process continues
hierarchically until reaching a size for $"m"$-qubit gates. Then these can be passed and handled by
third-party native synthesis tools.  The last stage of QFAST is
hierarchically 
reassembling the circuit using the native implementation of each
generic gate added in the first stage.

The three stage approach in QFAST enables composability and tunability
of approximations. First, we can plug in any native synthesis tool, at
any qubit concurrency, based on its perceived quality of solution or
execution speed. Second, we can decouple the ``distance'' metrics in
the generic gate space from the distance metrics in the native gate
space, thus enabling finer approximations of the input circuit.

QFAST has been evaluated on a series of algorithms used in previous
synthesis and circuit mapping studies~\cite{noisemap,bassman2020domainspecific,cowtan2019phase,davis2019heuristics,murali2019noise}.
We
compare its performance directly against the QSearch~\cite{qsearch} optimal
algorithm and against the IBM Qiskit synthesis~\cite{qiskit} modules. To
demonstrate composability, we show experiments when leveraging two-qubit
native optimal depth KAK decomposition, two- and three-qubit Qsearch.

The results indicate that QFAST behaves well in practice. When
compared against depth optimal synthesis in QSearch on circuits up to
four qubits, QFAST produces circuits on average $1.19\times$ longer 
with an average reduction in execution time of $3.55\times$. QFAST scales up to
seven qubits.  When compared against the sythesis algorithms deployed
in IBM Qiskit, QFAST produces circuits on average $156\times$ shorter,
albeit with an average time penalty of $15\times$. When compared
directly against traditional compilation approaches in IBM Qiskit, the
QFAST circuits are on average $10\times$ shorter. 

The rest of this paper is organized as follows.
In Section~\ref{sec:bg} we introduce the problem, motivation and
provide a short primer on quantum computng. In Section~\ref{sec:encoding}
we describe our circuit encoding used in the algorithm,
which is described in Section~\ref{sec:alg}.
We include an evaluation in Section~\ref{sec:eval} and a discussion
in Section~\ref{sec:disc}. Finally, we present related works in Section~\ref{sec:related}
and conclude in Section~\ref{sec:conc}

\section{Background}
\label{sec:bg}

In quantum computing, a qubit is the basic unit of quantum
information. Their general
quantum state is represented by a linear combination of two orthonormal basis states (basis vectors).  The most common basis is the equivalent
of the 0 and 1 values used for bits in classical information theory,
respectively {\footnotesize $\ket{0} = \left( \begin{array}{r} 1 \\ 0 \\ \end{array}
\right)$ } and {\footnotesize  $\ket{1} = \left( \begin{array}{r} 0 \\  1
                                                   \\  \end{array} \right)$}.

                                             The generic qubit
state is a superposition of the basis states, i.e. $\ket{\psi} =
\alpha \ket{0} + \beta \ket{1}$, with complex amplitudes $\alpha$ and
$\beta$ such that $|\alpha|^2+|\beta|^2=1$.

The prevalent model of quantum computation is the circuit model introduced
by Deutsch~\cite{qcircuit}, where information carried by qubits  (wires) is modified by
quantum gates, which mathematically correspond to unitary operations. A complex
square matrix U is {\bf unitary} if its conjugate transpose $U^*$ is  its
inverse, i.e. $UU^* = U^*U = I$.

In the circuit model, a single qubit gate is represented by a $2 \times 2$
unitary matrix U. The effect of the gate on the qubit state is obtained by
multiplying the U matrix with the vector representing the quantum
state $\ket{\psi'} = U\ket{\psi}$.

The most general form of the unitary for a single qubit
gate is  the ``continuous''
or  ``variational'' gate representation. {\footnotesize
        $U3(\theta,\phi,\lambda) = \left( \begin{array}{rr}
                                            cos{\frac{\theta}{2}}  & -e^{i\lambda}sin{\frac{\theta}{2}} \\
  e^{i\phi}sin{\frac{\theta}{2}} &
                                   e^{i\lambda+i\phi}cos{\frac{\theta}{2}}
                                            \\ \end{array} \right)$}

\begin{figure*}
    \centering
    \begin{subfigure}[b]{\textwidth}
        \centering
        \includegraphics[keepaspectratio=true,height=.75in]{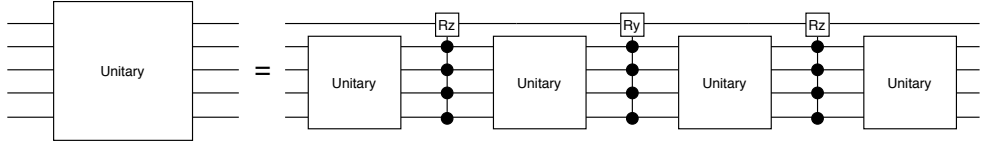}
        \caption{\label{fig:topdown} \it \footnotesize Top-down synthesizers follow prescribed, simple rules to decompose large unitaries into smaller ones while maintaining equality.}
        
    \end{subfigure}
    \begin{subfigure}[b]{\textwidth}
        \centering
        \includegraphics[keepaspectratio=true,height=.75in]{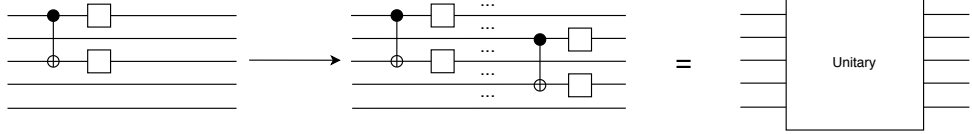}
        \caption{ \label{fig:bottomup} \it \footnotesize Bottom-up synthesizers start with an empty circuit and build up to equality.}
       
    \end{subfigure}
    \caption{\label{fig:synthesis} \it \footnotesize  Quantum synthesizers are either top-down or bottom-up.}
    
\end{figure*}

\begin{figure}
    \centering
    \begin{subfigure}[b]{\columnwidth}
      \centering
        \includegraphics[keepaspectratio=true,height=.75in]{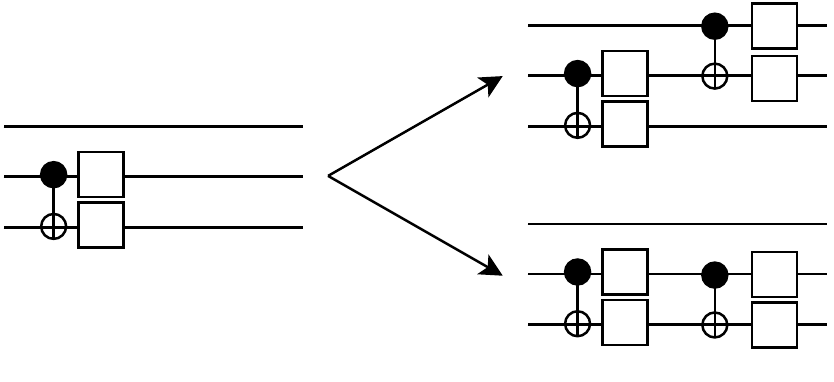}
        \caption{\label{fig:qsearchtree}\it \footnotesize  QSearch uses native gates in synthesis and searches for structure in their circuit space. Each node in their circuit space is a valid circuit structure that has edges to circuits deeper by a 2-qubit native gate.}
        
    \end{subfigure}
    \begin{subfigure}[b]{\columnwidth}
        \centering
        \includegraphics[keepaspectratio=true,height=.75in]{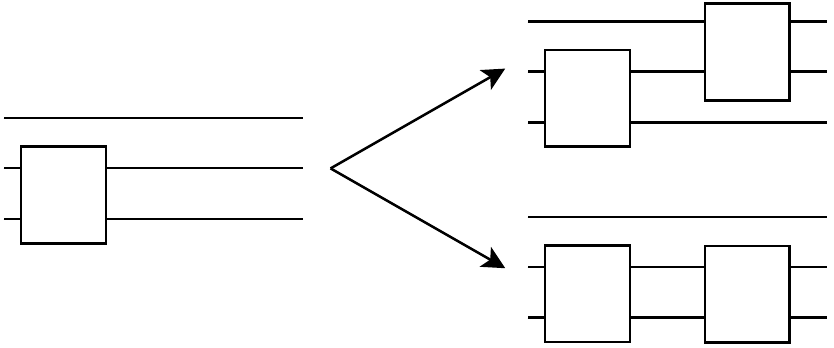}
        \caption{\label{fig:qfasttree} \it \footnotesize QFAST uses block unitary matrices of arbitrary size. In this figure, 2-qubit blocks are used and a similar tree is constructed.}
        
    \end{subfigure}
    \caption{\label{fig:trees} \it \footnotesize The 2-qubit unitary blocks are more expressive than one, fixed two-qubit gate followed by single-qubit rotations; Nodes 2 levels deep in the unitary block tree can only be expressed with nodes 6 levels deep in the native gate tree.}
    \end{figure}

A quantum transformation (algorithm, circuit)  on $n$-qubits is
represented by a unitary matrix U of size $2^n \times 2^n$.  A circuit is
described by an  evolution in space (application on qubits) and time of gates.
Figure~\ref{fig:synthesis} and \ref{fig:trees} shows a few examples of circuit.

\parah{Circuit Synthesis} The goal of circuit synthesis is to
decompose unitaries from $U(2^n)$ into a product of terms, where each
individual term  (e.g. from $U(2)$ and $U(4)$) captures the application of a
quantum gate on individual qubits.
The quality of a synthesis algorithm is evaluated by the number of gates in the
circuit it produces and by the distinguishability of the solution from the
original unitary.

Circuit length provides one of the main optimality criteria for
synthesis algorithms: shorter circuits are better. \cn count is a
direct indicator of overall circuit length, as the number of single
qubit generic gates  introduced in the circuit is proportional with a
constant given by decomposition (e.g. $ZXZXZ$) rules. As \cn gates
have low fidelity on NISQ devices, state-of-the-art
approaches~\cite{qsearch,raban,synthcsd} directly attempt to minimize their
count. Longer term, single qubit gate count is likely to augment the
quality metric for synthesis.

Synthesis algorithms use distance metrics  to assess the solution quality, and
their goal is to minimize $\|U_T - U_C\|$, where $U_T$ is the unitary
that describes the target transformation and $U_C$ is the computed
solution. They choose an error threshold $\epsilon$ and use it for
convergence, $\|U_T - U_C\| \le \epsilon$. Early synthesis algorithms use the diamond norm, while more recent efforts~\cite{qsearch,qaqc,HSnormsynth} use the Hilbert-Schmidt inner product between the conjugate transpose of $U_T$ and $U_C$.
This is motivated by its lower computational overhead.
\begin{equation} \label{eq:hsn}
   \langle U_T, U_C \rangle_{HS} = Tr(U_T^{\dag} U_C)
\end{equation}

\parah{Top-Down Synthesis}
These algorithms follow prescribed, simple rules to decompose large
unitaries into a tensor product of smaller terms or into a product of
symmetric matrices. Figure \ref{fig:topdown}
illustrates the Quantum Shannon Decomposition (QSD) \cite{shende2006synthesis}, which
breaks an $n$-qubit unitary into four  $(n-1)$-qubit
unitaries and three multi-controlled rotations. Like most top-down methods,
synthesis with QSD is  quick, but 
circuit depth grows exponentially. Overall, these techniques
are memory limited, rather than computationally limited. 

The only known depth optimal rule based algorithm is the
KAK-decomposition \cite{tucci2005kak}, which is valid only for two-qubit
operations.  Due to its optimality,  KAK  has been used in both
bottom-up and top-down compilers. For example, UniversalQ \cite{uq}
 implements  multiple top-down methods, some
exposed directly by IBM Qiskit. Their version
of QSD stops when reaching two-qubit blocks which
are instantiated to native gates by KAK.

\parah{Bottom-Up Synthesis} These algorithms, described in Figure \ref{fig:bottomup},
start with an empty circuit and attempt to place simple building
blocks until equality is formed.

QSearch~\cite{qsearch} introduces an optimal depth, topology aware synthesis
algorithm that has been demonstrated to be extensible across native
gate sets (e.g. \{$R_X, R_Z, CNOT$\},  \{$R_X, R_Z, SWAP$\}) and to
multi-level systems such as qutrits.
The approach employed in QSearch is canonical for the operation of
other synthesis approaches that employ numerical
optimization.

Conceptually, the synthesis problem can be thought as a
search over a tree of possible circuit structures. A search algorithm
provides a principled way to walk the tree and evaluate candidate
solutions. For each candidate, a numerical optimizer instantiates the
function (parameters) of each gate in order to minimize some distance
objective function.

QSearch works by extending the circuit structure a layer at a time. At each step
the algorithm places a 2-qubit expansion operator in all legal placements. For the CNOT gate set, the
operator contains one CNOT gate and two  $U3(\theta,\phi,\lambda)$ gates.
QSearch then evaluates these candidates using numerical optimization to
instantiate {\it all} the single qubit gates in the structure. An
A*~\cite{astar} heuristic determines which of the candidates is
selected for another layer expansion, as well as the destination of
backtracking steps. Figure~\ref{fig:qsearchtree} illustrates this process for
a three qubit circuit.

Although theoretically able to solve for any circuit size,
the scalability of QSearch is limited in practice to four qubit programs
due to several factors.
The
A* strategy determines the number of solutions evaluated: at best this
is linear in depth, at worst it is exponential. Our examination of
QSearch performance indicates that its scalability is limited to four
qubits first due to the presence of too many deep backtracking
chains. Any technique to reduce the number of candidates, especially
when deep, is likely to improve performance. 

As each
expansion operator has two single-qubit gates, accounting for six\footnote{In
  practice, QSearch uses 5 parameters due to commutativity rules
  between single qubit and
  CNOT gates.} parameters, circuit
paramaterization grows linearly with depth. Numerical
optimizers scale at best with a very high degree polynomial in
parameters, making optimization of long circuits challenging. 

\comment{To address these issues, the Qsearch team developed LEAP \cite{}, an extra set
of heuristics for the search, which greatly improved its scalability.
However, the nature of these types of bottom-up synthesizers eventually limits
LEAP at 5-6 qubits. This is for two reasons. First, the number of partial
solutions grows large quickly, especially with larger inputs. Each partial
solution is passed to an expensive numerical optimizer. Second, the cost of
each partial solution evaluation grows exponentially due to the larger
number of parameters to optimize. In a standard run where the building block
is composed of a single CNOT and two U3 gates, each block adds six parameters
to optimization problem.}

\parah{QFAST Approach} From the above discussion, several things
become apparent. Top-down methods scale to a larger number of qubits
than search based methods, and the quality of their solution can be
improved by the introduction of (nearly) optimal depth techniques that
work on more than two-qubits. The higher the number of qubits handled
by the bottom synthesis, probably the higher the impact on the quality
of the solution. Optimal search based techniques are
limited in scalability first by the search algorithm, second by the
scalability of numerical optimization. QFAST improves the
synthesis scalability while providing good solutions through very simple intuitive
principles:

\begin{enumerate}
    \item As small two-qubit building blocks may lack “computational power”,
		  we use generic blocks spanning a configurable number of qubits.
		  See Figure \ref{fig:qfasttree} for an example of the tree with
		  two-qubit building blocks. In this example, a
                    depth $two$ partial
		  solution could  express circuits that are up to
                  depth $six$\footnote{Any generic 2 qubit unitary
                    expands in at most three \cn gates, per KAK.}  in the
		  Qsearch tree.
	\item As the number of partial solutions and their evaluations may hamper
		  scalability, we conflate the numerical optimization and search
		  problem. We do this by using a continuous circuit space. At each step,
		  the circuit is expanded by one layer. Given an $n$-qubit circuit,
		  a layer encodes an arbitrary $m$-qubit operation on any $m$-qubits,
		  with $m < n$. Thus, our formulation does not having a branching
		  factor and solves combinatorially less optimization problems.
\end{enumerate}

\section{Gate and Circuit Representations}\label{sec:model}
\label{sec:encoding}

QFAST models a circuit as a sequence of parameterized gates.
Each gate has a function (hence a size), and a location.

The function
encodes the operation (quantum transformation) performed on an
associated number of qubits. 
When operating with a gate whose function is parameterized, we refer
to it as a {\it variable function gate}. Whenever gate parameterization
is numerically instantiated, the gate becomes {\it fixed function}.

The location describes the set of qubits
a gate is applied on, as placed in a larger circuit. A {\it variable
location} $m$-qubit gate is associated with a set of $n$-qubits, $n >
m$. In this case the gate can be applied to any valid subset of $m$-qubits
from the total $n$-qubits defined by the target topology. A {\it fixed location} $m$-qubit gate
operates on exactly $m$-qubits.

The QFAST algorithm uses two parameterizations:
{\it variable function with fixed location} gates, and {\it variable function
with 
variable location} gates.
The second parameterization allows us to conflate search and optimization.
In this section, we first describe how we encode gate variable function in a
fixed size gate. Then we build on the function encoding to encode
variable location.

\subsection{Encoding of Gate Function}
A gate's function is given by a unitary matrix.
As such, encoding gate function is equivalent to
structuring the unitary group.
Conveniently, the unitary group $U(2^n)$ is a Lie group.
It's Lie algebra $\mathfrak{u}(2^n)$ is the set of
$2^n \times 2^n$ skew-Hermitian matrices.
Using the Pauli group as generators for Hermitian matrices,
we can construct the unitary group in the following way:

$$U(2^n) = \{e^{\I(\vec{\alpha} \cdot \vec{\sigma^{\otimes n}})} \mid\ \vec{\alpha} \in \mathbb{R}^{4^n}\}$$
where $\vec{\sigma} = \{\sigma_i, \sigma_x, \sigma_y, \sigma_z\}$ are the Pauli matrices, and
$\vec{\sigma}^{\otimes n} = \{\sigma_j \otimes \sigma_k \mid \sigma_j \in \vec{\sigma}, \sigma_k \in \vec{\sigma}^{\otimes n-1} \}$
are the $n$-qubit Pauli strings.

This provides a useful parameterization of unitary operations on $n$-qubits.
We can then define an $n$-qubit gate's function with $2^n$ parameters as:
$$G(\vec{\alpha}) = e^{\I(\vec{\alpha} \cdot \vec{\sigma^{\otimes n}})}$$
This unitary-valued function is smooth and infinitely-differentiable.
Its derivative is given by the derivative of the exponential map \cite{rossmann2006lie}, but
when evaluating QFAST, we used the Padé approximation method with scaling and squaring \cite{branvcik2008matlab}
to compute the derivative.

\subsection{Encoding of Gate Location}

A gate's location determines which qubits it affects.
One simple way to encode a fixed location is to map
the Pauli strings that define the gate function to higher-order ones.

Given $Q$ a fixed $m$-qubit location on an $n$-qubit circuit
--- a $m$-length sequence of qubit indicies that are all less than $n$ ---
we define a map from $m$-qubit Pauli strings to $n$-qubit Pauli strings:

$$\pi_Q : \vec{\sigma}^{\otimes m} \longrightarrow \vec{\sigma}^{\otimes n}$$

This map inserts $n-m$ identities into the $m$-qubit Pauli string in positions
not specified in the location. For example, if we are given a 2-qubit location
$Q = (0, 1)$ on a 3-qubit circuit, then $\pi_Q(XX) = XXI$. If instead,
$Q = (0, 2)$, then $\pi_Q(XX) = XIX$.

This leads to a parameterization of an $m$-qubit gate with variable function
and a fixed location on an $n$-qubit circuit.

$$F(Q, \vec{\alpha}) = \exp({\I(\vec{\alpha} \cdot \pi_Q(\vec{\sigma^{\otimes m}}))})$$

If instead of a fixed location, we want variable location, given a set of valid locations,
we can simply multiplex all possible locations. For example, if we want a formulation
of a gate with variable function that affects either qubits $Q_0 = (0, 1)$ or qubits
$Q_1 = (1, 2)$, we simply write:

$$\exp({\I[ l_0(\vec{\alpha} \cdot \pi_{Q_0}(\vec{\sigma^{\otimes m}})) + l_1(\vec{\alpha} \cdot \pi_{Q_1}(\vec{\sigma^{\otimes m}})) ]})$$

Here either $l_0$ or $l_1$ is 1 and the other is 0. If $l_0$ is one, then
the formulation chooses the location given by $Q_0$. Likewise, if $l_1$ is one,
then the formulation chooses the location given by $Q_1$. This can be extended
to any number of possible locations $\vec{Q}$:

$$V(\vec{Q}, \vec{\alpha}, \vec{l}) = \exp({\I\sum\limits_{Q \in \vec{Q}}{l_Q \cdot \vec{\alpha} \cdot \pi_Q(\vec{\sigma^{\otimes m}})}})$$

\subsection{Direct Mapping of Pauli Strings}
Using the variable function with fixed location $F(Q, \vec{\alpha})$ and
the variable location and function $V(\vec{Q}, \vec{\alpha}, \vec{l})$
gates, it is enough to implement an algorithm that replaces search
with numerical optimization as shown in our first unpublished version of QFAST~\cite{younis2020qfast,Younis:EECS-2020-53}.

In  this formulation,  we solve a mixed integer-real optimization
problem, where the location and  the associated discrete $\vec{l}$
values are continuously approximated using the
{\tt softmax}~\cite{bishop2006pattern} function\footnote{The softmax function
  takes as input a vector $z$ of $K$ real numbers, and normalizes it
  into a probability distribution consisting of $K$ probabilities
  proportional to the exponentials of the input numbers. That is,
  prior to applying {\tt softmax}, some vector components could be
  negative, or greater than one; and might not sum to 1; but after
  applying  {\tt softmax}, each component will be in the interval $( 0 , 1 )$, and the components will add up to 1, so that they can be interpreted as probabilities. Furthermore, the larger input components will correspond to larger probabilities. }.

While the resulting implementation was able to produce good quality
circuits, its scalability and numerical stability proved more challenging. 
First, the matrices $\pi_Q(\vec{\sigma^{\otimes m}})$ are as large as the
full circuit unitary rather than the gate's unitary. Second, there are
$4^m$ of these unitaries. This latter point is worse in the variable location
model, where there are $|\vec{Q}|$ groups of $4^m$ full-sized
matrices.
The large dimensionality problem proved challenging to existing
optimization packages, thus the algorithm was numerically
unstable.

\begin{figure}
    \centering
	\includegraphics[width=\columnwidth]{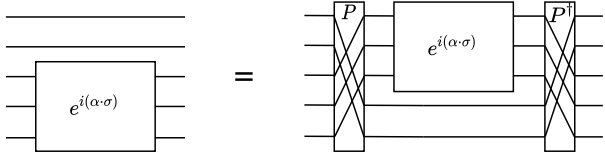}
    \caption{\label{fig:permutation} \it \footnotesize Permutations can be used to encode location.
			 A gate acting on the last 3-qubits of a 5-qubit circuit can be represented
			 by a gate acting on the first 3-qubits preceded and followed by a permutation.}
    
\end{figure}

\subsection{Permutations and Locations}\label{perms}

We can  sidestep these problems by using permutation matrices.
This is because there exists a permutation matrix $P_Q$ such that
$$F(Q, \vec{\alpha}) = P_Q (G(\vec{\alpha}) \otimes I) P_Q^T$$
This is illustrated in Figure \ref{fig:permutation}. Now $F(Q, \vec{\alpha})$
uses the gate's unitary in the equation and as such is much less costly
to represent. The size of the Pauli matrices are now gate-sized
rather than full-sized. This reduces the size of the model by an exponential factor.

Similarily, we can write the variable location model by multiplexing
the permutation matrices rather than groups of Pauli matrices:

$$V(\vec{Q}, \vec{\alpha}, \vec{l}) = (\sum\limits_{Q \in \vec{Q}}{l_Q \cdot P_Q}) (G(\vec{\alpha}) \otimes I) (\sum\limits_{Q \in \vec{Q}}{l_Q \cdot P_Q^T})$$

This formulation does use $|\vec{Q}|$ full-sized permutation matrices; however, since we do not need $|\vec{Q}|$ groups of pauli matrices, this
is exponentially less costly than the variable location encoding using 
mapped Pauli strings. The resulting complexity of the model given by the number of parameters is $4^m + |\vec{Q}|$.

\section{Synthesis Algorithm}
\label{sec:alg}

\begin{figure*}[htbp!]
    \centering
    \includegraphics[keepaspectratio=true,height=4in]{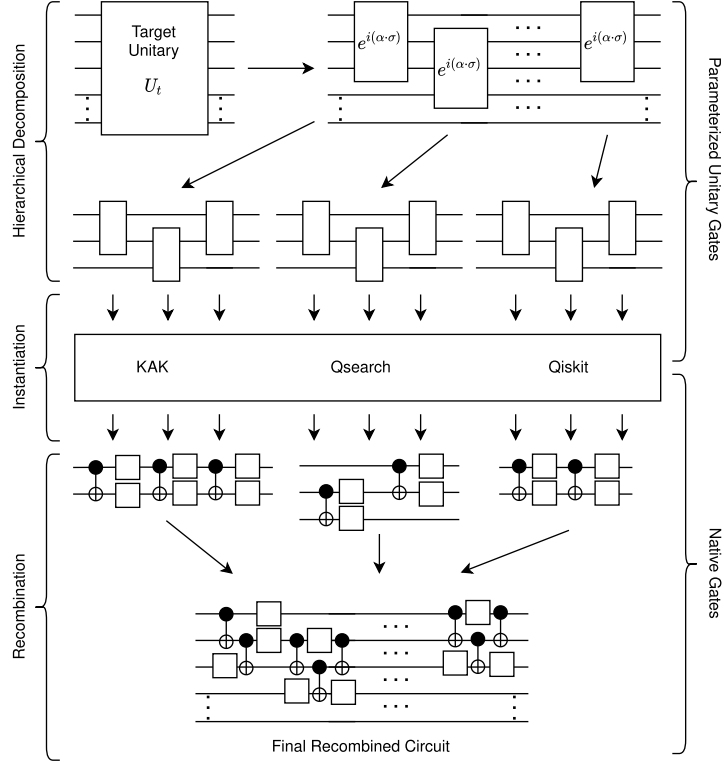}
	\caption{	\label{fig:qfast_alg}   \it \footnotesize QFAST is broken down into three phases. In the first phase,
			 decomposition, the target unitary is hierarchically broken down
			 into smaller blocks. Up until this point, the gates are
			 represented using the gate models described in section
			 \ref{sec:model}. During the next phase, instantiation, the
			 block unitary gates are converted into native gates with
			 third-party synthesis tools like Qsearch or with the
			 KAK decomposition. Finally, during recombination the circuit
			 is pieced back together and optimized where possible.}

\end{figure*}

QFAST's uses a hierarchical synthesis algorithm  split into three
stages, as described in 
Figure \ref{fig:qfast_alg}.
First, the target unitary is decomposed into parameterized unitary gates
 using the permutation model described in section \ref{perms} and the
 process described in Section~\ref{subsec:decomp}.
Next, the parameterized unitary gates are instantiated into native gates through
the use of a third-party synthesis tool, as described in
Section~\ref{subsec:inst}. As any native synthesis tool can be
selected, this composability allows us to tune the quality of the
solution together with the synthesis execution speed. 
Finally, the circuit is recombined into the final solution as
described in Section~\ref{subsec:recomb}.

\subsection{Decomposition}
\label{subsec:decomp}

\begin{figure}[htbp!]
	\centering
	\includegraphics[keepaspectratio=true,height=.75in]{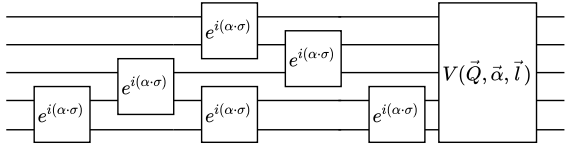}
	\caption{	\label{fig:qfast_decomp} \it \footnotesize
          Decomposition step in QFAST. Variable function with variable
        location gates are appended to a prefix circuit formed of
        variable function with fixed location gates. A step of
        numerical optimization will convert the newly added gate to a
        variable function with fixed location gate. } 

\end{figure}

Starting with an empty circuit, decomposition adds parameterized unitary gates
layer by layer until the circuit is close to the target unitary.
During construction, a partial solution contains only variable
function with fixed location gates. At each step, the algorithm
extends the partial solution  with a single
variable function with variable location gate, as illustrated in Figure~\ref{fig:qfast_decomp}. The new candidate
is then passed to a gradient-based optimizer, which solves for the parameters.
In doing so, the optimizer solves for the function of all gates
and the location of the ``head'' gate. As a result, the variable-location
gate has becomes a fixed-location gate. The
process repeats until convergence, specified as a threshold on the
distance between the target unitary and the partial solution.  

Numerical optimization can easily encounter plateaus and local minima due to the complexity of the
circuit and its associated objective function. During decomposition
this manifests as the new candidate solution having a larger distance
from target than the previous one.  QFAST alleviates this 
by  repeating  the same operation/optimization with a more restricted topology for the head gate.
In this case, the location that was chosen by the optimizer is removed
from the possible locations for the head gate. We perform this action
because chosing location seems harder for optimizers than solving only
for gate function.

Decomposition transforms a target unitary into a circuit of smaller-sized
unitaries. A benefit of using general unitary gates during decomposition,
is that the process can be applied recursively. This gives a hierarchical
decomposition algorithm.

\subsection{Instantiation}
\label{subsec:inst}

The decomposition stage produces a candidate circuit composed of
variable function blocks. While these can perform any computation, they are not directly
executable on hardware and we need a stage where blocks are transformed
and rewritten into hardware native gates.This is the stage where we
can freely leverage  previous approaches and plug in any third-party
synthesis tool. This composablility
gives QFAST portability across hardware architectures.

Although any solution can be used at this stage, the obviously mandatory
choices
are optimal depth topology aware algorithms: KAK~\cite{tucci2005kak}
decomposition for two-qubit blocks and QSearch for three-qubit blocks.

\subsection{Recombination}
\label{subsec:recomb}

Finally, the recombination stage will stitch together  into a complete
circuit  all the sub-circuits produced
during {\it instantiation}, each
associated with  a variable function gate. There is the opportunity to
apply circuit optimization techniques during this stage. During instantiation,
third-party synthesis tools are responsible for optimzing the native gate
sequence for each block. Recombination is responsible for putting the circuit
together, but there is also an opportunity here to optimize the native gate
sequence across block boundaries. We use traditional gate-level optimization
techniques to cancel or combine gates across block
boundaries.

\subsection{Distance Function}

The goal of synthesis is to generate a circuit that implements a unitary $U_C$
such that $||U_T^\dagger U_C - I|| \leq \epsilon$, where the $U_T$ is the target
unitary, $\epsilon$ is some small number, $I$ is the identity matrix. If
$\epsilon$ is 0, or equivalent to machine floating point epsilon, then the
synthesis is said to be exact, otherwise it is approximate.

The Frobenius norm $||U_T^\dagger U_C - I||$ is equivalent to $2 - 2Re(Tr(U_T^\dagger U_C))$.
The above equation can be simplified and scaled in the context of synthesis to
the following:

$$\Delta_F(U_C, U_T) = 1 - \frac{Re(Tr(U_T^{\dagger}U_C))}{d}$$

However, the $\Delta_F$ function is not tolerant of global-phase. Global-phase
is irrelevant in quantum computation because it does not affect measurement.
Being tolerant of global-phase implies that equivalent unitaries up to global
phase have zero distance. $\Delta_F$ can determine that two such unitaries
have arbitrary large distance. The function can be made global-phase tolerant:

$$\Delta(U_C, U_T) = 1 - \frac{|Tr(U_T^{\dagger}U_C)|}{d}$$

We use the above as our cost function during optimization. A value of zero
implies $U_C$ and $U_T$ are equivalent, and the largest value possible is one,
which implies the unitaries are very far apart.

\section{Evaluation}
\label{sec:eval}

\parah{Software Implementation} QFAST is implemented in Python 3 using
SciPy's implementation of
the L-BFGS~\cite{byrd1995limited}  optimization algorithm. The {\tt qfast} package is made available
as part of the BQSKit tool suite through PyPI \cite{pypi} and source code  can be found
at {\tt http://github.com/BQSKit/qfast}.  The software accepts as
input a target unitary matrix, and the output is presented to users as an OpenQASM 2.0 \cite{openqasm}
circuit/program. The following evaluation was performed
on   a desktop computer with a 3.3GHz AMD 2950X processor with 16 cores for a total of 32 threads.

\parah{Benchmarks}  The benchmark suite is composed of third party
algorithms used by other evaluation
studies~\cite{noisemap,bassman2020domainspecific,cowtan2019phase,davis2019heuristics,murali2019noise}. We consider
circuits with known optimal implementations, such as Quantum Fourier
Transform~\cite{qft}, HHL~\cite{hhl} and important quantum kernels
such as Toffoli gate, multiplier, adder etc. These allow us to gauge
the optimality of our solution.

We also consider circuits from domain generators such as the
Variational Quantum Eigensolver~\cite{McClean2015} (VQE) or Transverse
Field Ising Models~\cite{bassman2020domainspecific,
  tfimshin,tfimlb} (TFIM).  TFIM is an exponent of chemical simulations
using time dependent Hamiltonians. In this case, domain-specific compilers repeat a
fixed function ansatz for every timestep and circuit depth grows
linearily. Domain tools concentrate in reducing ansatz depth
and can't avoid linear growth. As no optimal depth circuit is known for these algorithms, they allow us to showcase the benefits of synthesis for circuit optimization. 

\parah{Comparison with State-of-the-Art} We evaluate QFAST alongside  state-of-art bottom-up and top-down
synthesis tools, as well as ``traditional'' compilers. To assess
optimality, we compare against the optimal depth topology aware
QSearch synthesis tool. To assess the quality of solution and
scalability for larger circuits we compare directly against the IBM
Qiskit~\cite{qiskit} synthesis passes. Note that these are based on the UniversalQ~\cite{uq}
top-down synthesis algorithms. Finally, we compare the
quality of the circuit generated by QFAST against the output of
traditional compilation and mapping with IBM Qiskit. 

\parah{Evaluation Metrics} We assess the
quality of the generated circuits along several metrics: 1) total \cn gate count; 2) total
count of single qubit gates (including $R_Z$ software gates); 3)
critical path length, the longest path of gates in the circuit; and 4) average gate parallelism.  The first
three metrics have been previoulsy used by synthesis and circuit
mapping studies. Average gate parallelism has been previously ignored
as a circuit quality metric due to severe scheduling restrictions
caused by QPU crosstalk. As our understanding of crosstalk increases,
we believe the presence of gate parallelism in circuits will become
important. For the purpose of this study, we use a simple measure 
defined as $\frac{Total\ Gate\ Count}{Critical\ Path\ Length}$.

Besides circuit structure related metrics, we discuss
the Hilbert-Schmidt distance of the solution and total  compilation time.

\parah{Experimental Results} We compare QFAST against state-of-the-art
tools for synthesis and circuit compilation.   To assess algorithm
optimality, we compare against the QSearch optimal depth topology aware
synthesis algorithm,  for circuits up to four
qubits.  We also compare against the IBM Qiskit synthesis algorithm
for all benchmarks. Another point of comparison is against traditional
compilation: in this case we compile and map any available input reference
circuits  directly with IBM Qiskit.

We evaluated  efficacy on two physical chip topologies: fully-connected
and linear (nearest-neighbor). Tables~\ref{tab:34q} and~\ref{tab:567q} show  metrics for all evalauted 
circuits, ranging from three to seven qubits. These demonstrate the
scalability of QFAST.  For each benchmark we
have have access to a reference circuit implementation obtained from
third parties, either domain generators or workloads considered by
other studies. The colums marked ``Mapped'' describe the metrics for
the reference inputs, when mapped by IBM Qiskit to a particular
topology. 

For circuits up to four qubits, we can compare directly against
QSearch, as shown in Table~\ref{tab:34q}. This allows us to evaluate
the optimality of QFAST.

\begin{table*}[!htp]\centering
    \scriptsize
    \makebox[\textwidth][c]{
        \setlength\tabcolsep{3pt}
        \begin{tabular}{|c|c|c|ccccccc|cccccccc|}\toprule
        \textbf{} &\textbf{} &\textbf{} &\multicolumn{7}{c}{\textbf{3 Qubits}} &\multicolumn{8}{c}{\textbf{4 Qubits}} \\\cmidrule{4-18}
        \textbf{} &\textbf{} &\textbf{} &\textbf{fredkin} &\textbf{toffoli} &\textbf{grover} &\textbf{hhl} &\textbf{or} &\textbf{peres} &\textbf{qft3} &\textbf{adder} &\textbf{vqe} &\textbf{tfim-4-1} &\textbf{tfim-4-10} &\textbf{tfim-4-22} &\textbf{tfim-4-60} &\textbf{tfim-4-80} &\textbf{tfim-4-95} \\\midrule
        \multirow{8}{*}{\textbf{CNOTs}} &\multirow{4}{*}{\textbf{All-to-All}} &\textbf{Qiskit Mapped} &8 &6 &7 &5 &6 &5 &6 &10 &76 &6 &60 &132 &360 &480 &570 \\
        & &\textbf{QFAST} &8 &8 &7 &3 &8 &7 &7 &15 &43 &8 &14 &16 &18 &14 &21 \\
        & &\textbf{QSearch} &8 &6 &7 &5 &7 &6 &7 &12 &22 &6 &12 &13 &12 &15 &12 \\
        & &\textbf{Qiskit Synthesized} &15 &9 &29 &13 &11 &11 &27 &66 &566 &124 &218 &218 &218 &218 &218 \\\cmidrule{2-18}
        &\multirow{4}{*}{\textbf{Linear}} &\textbf{Qiskit Mapped} &12 &13 &14 &11 &11 &9 &8 &20 &85 &6 &60 &132 &360 &480 &570 \\
        & &\textbf{QFAST} &8 &8 &7 &4 &8 &7 &8 &36 &40 &6 &10 &10 &12 &12 &23 \\
        & &\textbf{QSearch} &8 &8 &7 &3 &8 &7 &7 &14 &24 &7 &12 &13 &12 &13 &12 \\
        & &\textbf{Qiskit Synthesized} &30 &17 &74 &30 &19 &28 &70 &247 &2630 &477 &523 &523 &523 &523 &523 \\\cmidrule{1-18}
        \multirow{8}{*}{\textbf{U3s}} &\multirow{4}{*}{\textbf{All-to-All}} &\textbf{Qiskit Mapped} &10 &8 &17 &10 &9 &9 &11 &11 &86 &7 &70 &154 &420 &560 &665 \\
        & &\textbf{QFAST} &19 &19 &17 &9 &19 &17 &17 &34 &91 &20 &32 &36 &40 &32 &46 \\
        & &\textbf{QSearch} &19 &15 &16 &13 &17 &15 &17 &26 &49 &16 &28 &28 &28 &31 &28 \\
        & &\textbf{Qiskit Synthesized} &19 &11 &42 &17 &17 &12 &39 &88 &671 &160 &261 &261 &261 &261 &261 \\\cmidrule{2-18}
        &\multirow{4}{*}{\textbf{Linear}} &\textbf{Qiskit Mapped} &23 &22 &30 &22 &22 &20 &18 &37 &106 &7 &70 &154 &420 &560 &665 \\
        & &\textbf{QFAST} &19 &19 &17 &11 &19 &17 &19 &76 &84 &16 &24 &24 &28 &28 &50 \\
        & &\textbf{QSearch} &19 &19 &17 &9 &19 &17 &17 &32 &53 &18 &28 &30 &28 &30 &28 \\
        & &\textbf{Qiskit Synthesized} &50 &32 &126 &49 &37 &45 &120 &410 &4169 &785 &851 &851 &851 &850 &851 \\\cmidrule{1-18}
        \multirow{8}{*}{\textbf{Depth}} &\multirow{4}{*}{\textbf{All-to-All}} &\textbf{Qiskit Mapped} &11 &11 &16 &11 &8 &8 &12 &11 &116 &10 &73 &157 &423 &563 &668 \\
        & &\textbf{QFAST} &17 &17 &15 &7 &17 &15 &15 &21 &61 &9 &21 &29 &29 &29 &35 \\
        & &\textbf{QSearch} &17 &13 &15 &11 &15 &13 &15 &19 &39 &13 &21 &24 &19 &27 &21 \\
        & &\textbf{Qiskit Synthesized} &29 &17 &56 &26 &21 &19 &51 &121 &1062 &227 &421 &421 &421 &421 &421 \\\cmidrule{2-18}
        &\multirow{4}{*}{\textbf{Linear}} &\textbf{Qiskit Mapped} &23 &24 &29 &23 &21 &18 &17 &32 &136 &10 &73 &157 &423 &563 &668 \\
        & &\textbf{QFAST} &17 &17 &15 &9 &17 &15 &17 &63 &63 &9 &13 &21 &25 &21 &31 \\
        & &\textbf{QSearch} &17 &17 &15 &7 &17 &15 &15 &27 &41 &15 &23 &23 &25 &23 &21 \\
        & &\textbf{Qiskit Synthesized} &56 &34 &139 &55 &38 &51 &132 &390 &3949 &770 &852 &852 &852 &852 &852 \\\cmidrule{1-18}
        \multirow{8}{*}{\textbf{Parallelism}} &\multirow{4}{*}{\textbf{All-to-All}} &\textbf{Qiskit Mapped} &1.64 &1.27 &1.50 &1.36 &1.88 &1.75 &1.42 &1.91 &1.40 &1.30 &1.78 &1.82 &1.84 &1.85 &1.85 \\
        & &\textbf{QFAST} &1.59 &1.59 &1.60 &1.71 &1.59 &1.60 &1.60 &2.33 &2.20 &3.11 &2.19 &1.79 &2.00 &1.59 &1.91 \\
        & &\textbf{QSearch} &1.59 &1.62 &1.53 &1.64 &1.60 &1.62 &1.60 &2.00 &1.82 &1.69 &1.90 &1.71 &2.11 &1.70 &1.90 \\
        & &\textbf{Qiskit Synthesized} &1.17 &1.18 &1.27 &1.15 &1.33 &1.21 &1.29 &1.27 &1.16 &1.25 &1.14 &1.14 &1.14 &1.14 &1.14 \\\cmidrule{2-18}
        &\multirow{4}{*}{\textbf{Linear}} &\textbf{Qiskit Mapped} &1.52 &1.46 &1.52 &1.43 &1.57 &1.61 &1.53 &1.78 &1.40 &1.30 &1.78 &1.82 &1.84 &1.85 &1.85 \\
        & &\textbf{QFAST} &1.59 &1.59 &1.60 &1.67 &1.59 &1.60 &1.59 &1.78 &1.97 &2.44 &2.62 &1.62 &1.60 &1.90 &2.35 \\
        & &\textbf{QSearch} &1.59 &1.59 &1.60 &1.71 &1.59 &1.60 &1.60 &1.70 &1.88 &1.67 &1.74 &1.87 &1.60 &1.87 &1.90 \\
        & &\textbf{Qiskit Synthesized} &1.43 &1.44 &1.44 &1.44 &1.47 &1.43 &1.44 &1.68 &1.72 &1.64 &1.61 &1.61 &1.61 &1.61 &1.61 \\\cmidrule{1-18}
        \multirow{8}{*}{\textbf{Time (s)}} &\multirow{4}{*}{\textbf{All-to-All}} &\textbf{Qiskit Mapped} &0.04 &0.04 &0.05 &0.05 &0.04 &0.08 &0.04 &0.05 &0.36 &0.03 &0.20 &0.40 &1.00 &1.33 &1.67 \\
        & &\textbf{QFAST} &1.82 &1.77 &1.82 &0.23 &4.57 &0.54 &0.70 &7.71 &553.79 &1.29 &13.19 &12.26 &10.87 &6.12 &11.29 \\
        & &\textbf{QSearch} &2.99 &1.89 &1.84 &0.47 &1.01 &0.60 &0.98 &34.57 &2006.31 &10.56 &42.59 &16.41 &31.73 &30.71 &51.12 \\
        & &\textbf{Qiskit Synthesized} &0.26 &0.14 &0.86 &0.24 &0.22 &0.19 &0.60 &1.58 &12.10 &2.85 &3.36 &3.50 &3.37 &3.52 &3.32 \\\cmidrule{2-18}
        &\multirow{4}{*}{\textbf{Linear}} &\textbf{Qiskit Mapped} &0.17 &0.15 &0.17 &0.15 &0.16 &0.18 &0.13 &0.20 &1.04 &0.06 &0.32 &0.66 &1.82 &2.54 &2.93 \\
        & &\textbf{QFAST} &1.66 &1.64 &1.78 &0.41 &1.60 &1.25 &1.89 &16.25 &201.63 &0.64 &1.77 &1.85 &3.00 &2.81 &6.08 \\
        & &\textbf{QSearch} &2.42 &1.62 &1.42 &0.21 &1.52 &1.13 &0.72 &32.61 &765.19 &1.93 &57.15 &18.82 &9.54 &12.80 &11.24 \\
        & &\textbf{Qiskit Synthesized} &0.45 &0.28 &1.96 &0.46 &0.37 &0.39 &1.13 &4.07 &41.59 &7.55 &7.02 &8.25 &6.44 &8.47 &7.00 \\
        \bottomrule
        \end{tabular}
    }
    \caption{Results for 3-4 qubit synthesis benchmarks.}
    \label{tab:34q}
\end{table*}

\begin{table*}[!htp]\centering
\tiny
\makebox[\textwidth][c]{
    \setlength\tabcolsep{3pt}
    \begin{tabular}{|c|c|c|cccccccccc|ccccc|ccc|}\toprule
        \textbf{} &\textbf{} &\textbf{} &\multicolumn{10}{c}{\textbf{5 Qubits}} &\multicolumn{5}{c}{\textbf{6 Qubits}} &\multicolumn{3}{c}{\textbf{7 Qubits}} \\\cmidrule{4-21}
        \textbf{} &\textbf{} &\textbf{} &\textbf{grover5} &\textbf{hlf} &\textbf{mul} &\textbf{qaoa} &\textbf{qft5} &\textbf{tfim-5-10} &\textbf{tfim-5-40} &\textbf{tfim-5-60} &\textbf{tfim-5-80} &\textbf{tfim-5-100} &\textbf{tfim-6-1} &\textbf{tfim-6-10} &\textbf{tfim-6-24} &\textbf{tfim-6-31} &\textbf{tfim-6-51} &\textbf{tfim-7-20} &\textbf{tfim-7-40} &\textbf{tfim-7-100} \\\midrule
        \multirow{6}{*}{\textbf{CNOTs}} &\multirow{3}{*}{\textbf{All-to-All}} &\textbf{Qiskit Mapped} &48 &13 &17 &20 &20 &80 &320 &480 &640 &800 &10 &100 &240 &310 &510 &240 &480 &1200 \\
        & &\textbf{QFAST} &70 &13 &18 &39 &46 &20 &20 &24 &22 &26 &12 &29 &26 &24 &28 &41 &* &* \\
        & &\textbf{Qiskit Synthesized} &570 &870 &77 &750 &580 &1025 &1025 &1025 &1025 &1025 &4006 &4474 &4474 &4474 &4474 &18653 &18653 &18653 \\\cmidrule{2-21}
        &\multirow{3}{*}{\textbf{Linear}} &\textbf{Qiskit Mapped} &131 &23 &22 &55 &31 &80 &320 &480 &640 &800 &10 &100 &240 &310 &510 &240 &480 &1200 \\
        & &\textbf{QFAST} &60 &55 &58 &69 &114 &12 &18 &20 &20 &21 &10 &16 &20 &22 &32 &24 &31 &40 \\
        & &\textbf{Qiskit Synthesized} &2503 &2578 &760 &2692 &2622 &2791 &2791 &2791 &2791 &2791 &13155 &13365 &13365 &13365 &13365 &58316 &58316 &58316 \\\cmidrule{1-21}
        \multirow{6}{*}{\textbf{U3s}} &\multirow{3}{*}{\textbf{All-to-All}} &\textbf{Qiskit Mapped} &78 &8 &16 &20 &29 &90 &360 &540 &720 &900 &11 &110 &264 &341 &561 &260 &520 &1300 \\
        & &\textbf{QFAST} &145 &31 &41 &83 &97 &45 &45 &53 &49 &57 &30 &64 &58 &54 &62 &89 &* &* \\
        & &\textbf{Qiskit Synthesized} &672 &976 &87 &861 &687 &1140 &1140 &1140 &1140 &1140 &4294 &4765 &4765 &4765 &4765 &19360 &19360 &19360 \\\cmidrule{2-21}
        &\multirow{3}{*}{\textbf{Linear}} &\textbf{Qiskit Mapped} &235 &37 &40 &93 &63 &90 &360 &540 &720 &900 &11 &110 &264 &341 &561 &260 &520 &1300 \\
        & &\textbf{QFAST} &125 &115 &121 &143 &233 &29 &41 &45 &45 &47 &26 &38 &46 &50 &70 &55 &69 &87 \\
        & &\textbf{Qiskit Synthesized} &4008 &4046 &1190 &4264 &4165 &4400 &4401 &4401 &4401 &4400 &20375 &20659 &20658 &20656 &20658 &89478 &89480 &89483 \\\cmidrule{1-21}
        \multirow{6}{*}{\textbf{Depth}} &\multirow{3}{*}{\textbf{All-to-All}} &\textbf{Qiskit Mapped} &85 &16 &26 &32 &26 &76 &286 &426 &566 &706 &16 &79 &177 &226 &366 &152 &292 &712 \\
        & &\textbf{QFAST} &123 &21 &33 &65 &85 &31 &33 &49 &29 &39 &13 &47 &29 &29 &33 &47 &* &* \\
        & &\textbf{Qiskit Synthesized} &1064 &1662 &138 &1451 &1089 &2008 &2008 &2008 &2008 &2008 &7872 &8841 &8841 &8841 &8841 &37048 &37048 &37048 \\\cmidrule{2-21}
        &\multirow{3}{*}{\textbf{Linear}} &\textbf{Qiskit Mapped} &200 &34 &40 &76 &44 &76 &286 &426 &566 &706 &16 &79 &177 &226 &366 &152 &292 &712 \\
        & &\textbf{QFAST} &99 &87 &77 &83 &151 &17 &21 &29 &21 &27 &13 &17 &25 &21 &45 &23 &31 &49 \\
        & &\textbf{Qiskit Synthesized} &3799 &3933 &1115 &4061 &3924 &4236 &4236 &4236 &4236 &4236 &19074 &19495 &19495 &19494 &19494 &82915 &82915 &82915 \\\cmidrule{1-21}
        \multirow{6}{*}{\textbf{Parallelism}} &\multirow{3}{*}{\textbf{All-to-All}} &\textbf{Qiskit Mapped} &1.48 &1.31 &1.27 &1.25 &1.88 &2.24 &2.38 &2.39 &2.40 &2.41 &1.31 &2.66 &2.85 &2.88 &2.93 &3.29 &3.42 &3.51 \\
        & &\textbf{QFAST} &1.75 &2.10 &1.79 &1.88 &1.68 &2.10 &1.97 &1.57 &2.45 &2.13 &3.23 &1.98 &2.90 &2.69 &2.73 &2.77 &* &* \\
        & &\textbf{Qiskit Synthesized} &1.17 &1.11 &1.19 &1.11 &1.16 &1.08 &1.08 &1.08 &1.08 &1.08 &1.05 &1.05 &1.05 &1.05 &1.05 &1.03 &1.03 &1.03 \\\cmidrule{2-21}
        &\multirow{3}{*}{\textbf{Linear}} &\textbf{Qiskit Mapped} &1.83 &1.76 &1.55 &1.95 &2.14 &2.24 &2.38 &2.39 &2.40 &2.41 &1.31 &2.66 &2.85 &2.88 &2.93 &3.29 &3.42 &3.51 \\
        & &\textbf{QFAST} &1.87 &1.95 &2.32 &2.55 &2.30 &2.41 &2.81 &2.24 &3.10 &2.52 &2.77 &3.18 &2.64 &3.43 &2.27 &3.43 &3.23 &2.59 \\
        & &\textbf{Qiskit Synthesized} &1.71 &1.68 &1.75 &1.71 &1.73 &1.70 &1.70 &1.70 &1.70 &1.70 &1.76 &1.75 &1.75 &1.75 &1.75 &1.78 &1.78 &1.78 \\\cmidrule{1-21}
        \multirow{6}{*}{\textbf{Time (s)}} &\multirow{3}{*}{\textbf{All-to-All}} &\textbf{Qiskit Mapped} &0.16 &0.05 &0.07 &0.07 &0.11 &0.22 &0.88 &1.19 &1.68 &2.03 &0.04 &0.28 &0.62 &0.80 &1.41 &0.61 &1.28 &3.16 \\
        & &\textbf{QFAST} &3187.40 &27.70 &86.79 &249.15 &499.49 &79.86 &69.38 &71.98 &77.42 &215.13 &23.14 &618.43 &191.99 &270.70 &684.63 &13222.11 &* &* \\
        & &\textbf{Qiskit Synthesized} &11.61 &14.50 &2.65 &14.61 &14.43 &14.35 &15.04 &14.59 &14.27 &16.52 &82.16 &62.93 &64.10 &63.34 &64.62 &307.57 &290.26 &286.75 \\\cmidrule{2-21}
        &\multirow{3}{*}{\textbf{Linear}} &\textbf{Qiskit Mapped} &1.12 &0.24 &0.38 &0.46 &0.34 &0.43 &1.75 &2.57 &3.39 &4.31 &0.09 &0.51 &1.30 &1.61 &2.60 &1.17 &2.41 &6.09 \\
        & &\textbf{QFAST} &992.38 &228.55 &213.94 &365.15 &1901.26 &7.67 &22.78 &26.63 &30.28 &21.01 &5.25 &61.68 &82.52 &408.35 &772.39 &2524.35 &2659.28 &8208.44 \\
        & &\textbf{Qiskit Synthesized} &33.20 &34.42 &12.38 &36.25 &38.37 &35.93 &35.53 &32.27 &34.11 &32.41 &170.08 &161.25 &156.66 &161.30 &159.81 &676.79 &676.95 &705.01 \\
        \bottomrule
        \end{tabular}
}
\caption{Results for 5-7 qubit synthesis benchmarks. (* implies the program timed-out after 12 hours.)}
\label{tab:567q}
\end{table*}

\begin{table*}[!htp]\centering
\scriptsize
\makebox[\textwidth][c]{
    \begin{tabular}{|c|c|c|ccccc|ccc|}\toprule
        \textbf{} &\textbf{} &\textbf{} &\multicolumn{5}{c}{\textbf{6 Qubits}} &\multicolumn{3}{c}{\textbf{7 Qubits}} \\\cmidrule{4-11}
        \textbf{} &\textbf{} &\textbf{} &\textbf{tfim-6-1} &\textbf{tfim-6-10} &\textbf{tfim-6-24} &\textbf{tfim-6-31} &\textbf{tfim-6-51} &\textbf{tfim-7-20} &\textbf{tfim-7-40} &\textbf{tfim-7-100} \\\midrule
        \multirow{4}{*}{\textbf{CNOTs}} &\multirow{2}{*}{\textbf{All-to-All}} &\textbf{2-Qubit Gates} &12 &29 &26 &24 &28 &41 &* &* \\
        & &\textbf{3-Qubit Gates} &38 &47 &82 &68 &84 &94 &127 &169 \\\cmidrule{2-11}
        &\multirow{2}{*}{\textbf{Linear}} &\textbf{2-Qubit Gates} &10 &16 &20 &22 &32 &24 &31 &40 \\
        & &\textbf{3-Qubit Gates} &16 &28 &30 &38 &62 &46 &51 &64 \\\cmidrule{1-11}
        \multirow{4}{*}{\textbf{U3s}} &\multirow{2}{*}{\textbf{All-to-All}} &\textbf{2-Qubit Gates} &30 &64 &58 &54 &62 &89 &* &* \\
        & &\textbf{3-Qubit Gates} &82 &100 &170 &142 &174 &195 &261 &345 \\\cmidrule{2-11}
        &\multirow{2}{*}{\textbf{Linear}} &\textbf{2-Qubit Gates} &26 &38 &46 &50 &70 &55 &69 &87 \\
        & &\textbf{3-Qubit Gates} &38 &62 &66 &82 &130 &99 &109 &135 \\\cmidrule{1-11}
        \multirow{4}{*}{\textbf{Depth}} &\multirow{2}{*}{\textbf{All-to-All}} &\textbf{2-Qubit Gates} &13 &47 &29 &29 &33 &47 &* &* \\
        & &\textbf{3-Qubit Gates} &57 &93 &161 &119 &137 &159 &223 &273 \\\cmidrule{2-11}
        &\multirow{2}{*}{\textbf{Linear}} &\textbf{2-Qubit Gates} &13 &17 &25 &21 &45 &23 &31 &49 \\
        & &\textbf{3-Qubit Gates} &25 &57 &43 &59 &97 &75 &75 &93 \\\cmidrule{1-11}
        \multirow{4}{*}{\textbf{Parallelism}} &\multirow{2}{*}{\textbf{All-to-All}} &\textbf{2-Qubit Gates} &3.23 &1.98 &2.90 &2.69 &2.73 &2.77 &* &* \\
        & &\textbf{3-Qubit Gates} &2.11 &1.58 &1.57 &1.76 &1.88 &1.82 &1.74 &1.88 \\\cmidrule{2-11}
        &\multirow{2}{*}{\textbf{Linear}} &\textbf{2-Qubit Gates} &2.77 &3.18 &2.64 &3.43 &2.27 &3.43 &3.23 &2.59 \\
        & &\textbf{3-Qubit Gates} &2.16 &1.58 &2.23 &2.03 &1.98 &1.93 &2.13 &2.14 \\\cmidrule{1-11}
        \multirow{4}{*}{\textbf{Time (s)}} &\multirow{2}{*}{\textbf{All-to-All}} &\textbf{2-Qubit Gates} &23.14 &618.43 &191.99 &270.70 &684.63 &13222.11 &* &* \\
        & &\textbf{3-Qubit Gates} &90.41 &1250.79 &2874.59 &1097.66 &529.49 &26667.53 &39124.92 &31699.46 \\\cmidrule{2-11}
        &\multirow{2}{*}{\textbf{Linear}} &\textbf{2-Qubit Gates} &5.25 &61.68 &82.52 &408.35 &772.39 &2524.35 &2659.28 &8208.44 \\
        & &\textbf{3-Qubit Gates} &13.30 &293.28 &103.57 &722.94 &1153.03 &4056.87 &911.26 &2431.82 \\
        \bottomrule
        \end{tabular}
}
\caption{A comparison between QFAST configured with 2-qubit blocks and 3-qubit blocks. (* implies the program timed-out after 12 hours.)}
\label{tab:blocks}
\end{table*}

\subsection{Circuit Depth}  

QFAST produces good quality
circuits. When compared against traditional compilation and mapping
with IBM Qiskit, QFAST produces circuits with an average of $10\times$ fewer \cn
gates and $5.2\times$ fewer $U3$ gates. These are reflected into an average $5.7\times$
decrease of the circuit critical path and an $1.03\times$ better
parallelism. When compared against IBM Qiskit synthesis, the resulting
circuit critical path is $156\times$ shorter and parallelism is
$2\times$ higher.

When compared against QSearch for circuits up to four qubits, although
not optimal, QFAST produces good quality circuits: the critical path
is $1.19\times$ longer, while the parallelism is $1.09\times$
higher. Note that QFAST is on average $3.55\times$ faster than
QSearch.

QFAST handles topology very well. The ratio between the average
critical path length when synthesizing  to a linear topology and to a fully
connected topology is $1.07\times$. The QSearch ratio is
$1.06\times$.

Note that QFAST and QSearch are particularly useful in optimizing  circuits such
as TFIM, where no optimal implementation is apriori known and the
circuit grows linearily with each time step simulation of the target
system/Hamiltonian. In the result tables, we indicate the number of
qubits and the associated time step, e.g. {\tt tfim-6-10}.

\subsection{Scalability and Execution Time}

For brevity, we only summarize the excution time trends, and 
note that QFAST scales up to seven qubits. QFAST is roughly $15\times$
slower than IBM Qiskit and $3.55\times$ faster than QSearch.  The
maximum compilation time observed was 220 minutes for a seven qubit
circuit.

An analysis indicates that $>90\%$ of the execution
time is spent in the {\it decomposition} stage, with rest spent in the
{\it instantiation stage} for synthesizing small blocks. Overall
$>99\%$ of the total execution time is spent in numerical
optimization.

The decision to conflate search and numerical optimization pays off in
terms of improved scalability, albeit at a small loss in circuit
quality. Scalability comes from  exploring a reduced number of partial
solutions. For example, for {\tt tfim-4-95}, the partial solution space for
QSearch contains 1,594,323 circuits and the algorithm explores 30,460
circuits. In contrast, for the same input problem, QFAST explores 17 partial
solutions, respectively.

\subsection{Impact of Native Back-end Synthesis}

All results in  Tables~\ref{tab:34q} and \ref{tab:567q} have been obtained with QFAST
decomposing to two-qubit generic gates, each instantiated with KAK as
a native synthesis back-end. For brevity, we note that results
obtained when replacing KAK with QSearch are identical in terms of
circuit quality and execution time.

Composing with QSearch allows us to study the sensitivity to the
granularity of decomposition into two- or three-qubit generic
blocks. Detailed results are presented in Table~\ref{tab:blocks}, and
summarized for six and seven-qubit circuits in Table~\ref{tab:bs}.
Overall, directly  targeting the smallest block possible generates
shorter circuits. The reason is best understood when examining the
results for Linear mapping. In this case, we observe a reduction from
94 two-qubit generic blocks to 45  three-qubit blocks required to
represent all circuits. This is expected, as three-qubit blocks have
higher ``computational'' power than two-qubit blocks. On the other
hand, the three-qubit blocks seem to have too much computational power
in this case, 
as they expand into long subcircuits. Their depth expansion factor
(seven \cn on average) when
instantiated into native gates, is not offset by the reduction in the
total number of blocks.

\begin{table}[!htp]\centering

\begin{tabular}{|c|c|c|c|c|}
  \hline
     &   \multicolumn{2}{c|}{All-to-All} &
                                           \multicolumn{2}{c|}{Linear}
                                           \\
\hline
     & B=2 & B=3 & B=2 & B=3 \\
\hline 
  CNOT & 160 & 709 &195 & 335 \\
     Total Blocks & 76 & 81 & 94 & 45 \\
      Avg. Bl.  Len & 2.1 & 8.7 & 2.07 & 7.4 \\
 \hline
     \end{tabular}

\caption{\label{tab:bs} \it \footnotesize Summary of  QFAST configured with 2-qubit
  blocks and 3-qubit blocks, when compiling six- and seven-qubit benchmarks.}

\end{table}

\subsection{Validation of Generated Circuits}

At synthesis time, the solution quality is assessed using
modifications on the Hilbert-Schmidt distance.  All solutions
evaluated produce circuits that are some small distance away
from the input unitary. Qiskit's synthesis is
closest to target with an average distance of $10^{-14}$, followed by
QSearch with an average of $10^{-12}$ and by QFAST with $10^{-6}$.

To test the quality of the circuits we have run simulations with inputs
set to all the standard basis state vectors and 1000 random state vectors.
For all circuits with a distance less than $10^{-3}$, the average output
state fidelity is greater than 0.9999 with ULP difference of $10^{-5}$.
The average error of two-qubit gates is on the order
of $10^{-2}$ \cite{PhysRevA.87.030301}, which leds us to believe the threshold chosen for  QFAST
is fine
for the NISQ era.

In the case of QFAST and QSearch, the precision threshold is
configurable. For the QSearch results we asked for $10^{-6}$
and obtained $10^{-12}$.
The QFAST results have been obtained  with $10^{-3}$ threshold for the
decomposition stage, while the back-end synthesis with KAK is
precise.  If desired, by increasing the threshold during decomposition, users  can
obtain even better solutions at the expense of running time.

\section{Discussion}
\label{sec:disc}

QFAST improves upon previous synthesis techniques
in either quality of solution or scalability: 1) it is much faster than optimal techniques with little loss in circuit quality; and 2) generates much better circuits than fast techniques, with low enough time penalty.
Overall, we find the QFAST results encouraging for the future practical
use of synthesis in quantum algorithm exploration and optimization in the NISQ era and beyond.

Scalability is always a concern for synthesis algorithms. The current implementation is not parallelized, besides any internal multi-threading parallelism that may be present in numerical optimizers. Most of the time is currently spent in the {\it decomposition} stage, which has no intrinsic parallelism. The little time spent in {\it instantiation} and {\it recombination} can be reduced, as these are embarassingly parallel steps on generic gates.  For QFAST to improve, advances in the speed of  numerical optimization are required, and we are already contemplating GPU based implementations.   

A more subtle problem is trying to determine the useful upper bound on circuit size in qubits which synthesis algorithms should handle. {\it  Is seven qubits good enough to use beyond the NISQ era?} Our recent work and QGo algorithm~\cite{wu2020optimizing}, which  combines circuit partitioning techniques with synthesis indicates that a practical upper bound may be lower than expected, as we show great depth reduction for circuits up to 100 qubits when using QSearch optimal synthesis on four-qubit blocks. In this context, the expectation is that deploying QFAST will reduce time to solution while improving final solution quality. Futhermore, we expect gradually lower return-on-investments when scaling the bottom algorithms up to more qubits.
An indication of the dynamics at work can be seen in the analysis of QFAST behavior with block size.

Examining the behavior of these synthesis algorithms with respect to topology raises some interesting conjectures to help guide users in their choice of the appropriate tool for a given problem. The determining factor seems to be the relationship between the topology and qubit interactions required on the original quantum algorithm side (e.g. QFT, adder), and the topology on the physical QPU side. For bottom-up search based techniques (QSearch), the quality of the solution is not affected by any mismatch, just the time to solution. In this case mapping a low logical  connectivity algorithm, such as TFIM, on a rich all-to-all physical connectivity QPU will increase the total time to solution by a large margin.  For QFAST, both the quality of the solution and the time to solution are affected. For example, we see a non-intuitive depth inversion  when mapping TFIM to an all-to-all physical topology: the solution circuits are longer than when mapping directly to a linear topology. Overall,  the results indicate that in practice, when using bottom-up approaches it is probably a good heuristic to map to the most restricted linear topology, irrespective of the physical chip topology.
The conjecture is reversed for top-down approaches which seem really challenged by anything other than all-to-all connectivity.

\section{Related Work}
\label{sec:related}

A foundational result  is provided by the
Solovay Kitaev (SK) theorem which 
relates circuit depth to the quality of the approximation~\cite{DawsonNielson05,Nagy16,ola15}. Different approaches~\cite{DawsonNielson05,ZXZ16,BocharovPRL12,MIM13,Qcompile16,ctmq,23gates,householderQ,CSD04,amy16,seroussi80}  have been
introduced since, with the goal of generating shorter depth circuits. 
These can be coarsely classified based on several
criteria: 1) target gate set; 2) algorithmic approach; and 3) solution distinguishability.

\parah{Target Gate Set}
Some algorithms target gates likely
to be used only when fault tolerant quantum computing materializes. Examples include
synthesis of z-rotation unitaries with
Clifford+V approximation~\cite{Ross15} or Clifford+T
gates~\cite{KMM16,KSV02,Paetznick2014}.
While these efforts propelled the field of synthesis, they are not 
used on NISQ devices, which offer a different gate set
(e.g. $U_3, R_x, R_z,CNOT$ and M\o lmer-S\o rensen all-to-all).
Several~\cite{raban,synthcsd,ionsynth,davis2019heuristics}  
algorithms, discussed below target these gates directly. From our perspective,
since QFAST is composable and can invoke any synthesizer for instantiation, the existence of these algorithms indicates
that QFAST is portable across gate sets. 

\comment{ For example, the z-rotation unitaries can be
synthesized with Clifford+V approximation~\cite{Ross15} or  with Clifford+T gates~\cite{KMM16}. The set of Pauli, Hadamard, Phase, CNOT 
gates form what is known as the Clifford group gates. When augmented with the T gate defined
as 
\[
T = \left(
\begin{array}{cc}
1 & 0 \\
0 & \zeta_8 
\end{array}
\right), \ \ \mbox{where} \ \ \zeta_8 = e^{i\pi/4},
\]
the gate set is universal.
t lead to better complexity $\mathcal{O}(\log^{1.75}(1/\epsilon))$ compared 
to the SK Algorithm.  \comment{$\mathcal{O}(\log(1/\epsilon))$ T-count scaling.}
}  

\parah{Algorithmic Approaches}
Most  earlier attempts inspired by
Solovay Kitaev use a recursive (or divide-and-conquer) 
formulation. More recent search based approaches are illustrated by the
Meet-in-the-Middle~\cite{MIM13}  algorithm.
Several  approaches \cite{23gates,householderQ} use techniques from linear algebra for
unitary/tensor decomposition, but there are open questions as to the suitability
for hardware implementation because  algorithms are expressed 
in terms of row and column updates of a matrix rather than in terms of qubits.

The state-of-the-art upper bounds on circuit depth are provided by
techniques~\cite{synthcsd,raban} that use Cosine-Sine
decomposition. The Cosine-Sine decomposition was first
used by~\cite{tucci}  for compilation purposes. In practice,
commercial compilers ubiquitously deploy only 
KAK decompositions for two qubit unitaries.
Khaneja and Glaser have applied the KAK Decomposition to more than
just 2-qubit systems \cite{khaneja2000cartan}.
For a 3-qubit system, it originally required 64 CNOTs \cite{vartiainen2004efficient}, which was later reduced to 40 CNOTs \cite{vatan2004realization}. We have shown above that this can be beaten by any of the three synthesis tools tested in this work.
IBM Qiskit (based in UniversalQ~\cite{raban}) is
an exponent evaluated in this paper.
The basic formulation of these techniques is topology
independent.
The published approaches are hard to extend to different qubit gate
sets.

\comment{
There are not many studies published about synthesis of qutrit based
circuits and qutrit gate sets.~\cite{qtsynth} describes a method
using Givens rotations and Householder decomposition. As techniques
for qubit based systems using a similar approach have been
proposed~\cite{23gates}, they may allow an easier combination of
qutrit and qubit based synthesis. }

Several techniques~\cite{ionsynth,qaqc,davis2019heuristics} use numerical optimization
and report results for systems with at most four qubits. They
describe the single qubit gates in their variational/continuous representation and
use optimizers and search to find a gate decomposition and
instantiation. From these, we compare directly against QSearch~\cite{davis2019heuristics} which
is the only published optimal and topology-aware technique. For our
purposes, all these techniques seem to solve a combinatorial number of
hard (low distance) optimization problems. We expect QFAST to scale
better while providing comparable results. Furthermore, due to its
composability, we can directly leverage any of these implementations.

Topology awareness is important for synthesis algorithms, with
opposing trends. Most formulations assume all-to-all connectivity.
Specializing for topology in linear algebra
decomposition techniques seems to increase circuit depth by
rather large constants, ~\cite{synthcsd} mention a factor of nine,
improved by~\cite{raban} to $4\times$.  Specializing for topology in
search and numerical optimization techniques seems to reduce circuit
depth and Davis et al~\cite{davis2019heuristics} report up to  $4\times$  reductions.
QFAST  behaves like the latter.

\parah{Solution Distinguishability} 
Synthesis algorithms are classified as exact or approximate based on
distinguishability.  This is a subtle classification criteria, as most
algorithms can be viewed as either.  For example,~\cite{MIM13}
proposed a divide-and-conquer algorithm called Meet-in-the-Middle
(MIM). Designed for exact circuit synthesis, the algorithm
may also be used to construct an $\epsilon$-approximate circuit. The results seem to indicate that the algorithm failed
to synthesize a three qubit QFT circuit. 

Furthermore, on NISQ devices, the target gate set of the algorithm
(e.g. T gate) may
be itself implemented as an approximation when using native gates.

We classify our approach as approximate since we accept solutions at a small distance from the original
unitary. In a sense, when algorithms move from design to
implementation, all become approximate due to numerical
floating point errors.

\comment{
It allows one to search for
circuits of depth $l$ by only generating circuits of depth at most $\lceil l/2 \rceil$ at the complexity of $\mathcal{O}(|\mathcal{V}_{n,\mathcal{G}}|^{\lceil l/2\rceil}\log |\mathcal{V}_{n,\mathcal{G}}|^{\lceil l/2 \rceil})$, 
where $\mathcal{V}_{n,\mathcal{G}}$ denotes the set of unitaries for depth-one 
$n$-qubit circuit. The MIM algorithm is flexible and allows weights to be
added to the gate set to account for the possibility that some gates,
such as those that do not belong to the Clifford group, may be more expensive
to implement. It also allows ancillas to be used in the synthesis.  The 
algorithm uses a number of heuristics to prune the search tree.  Although
it was originally designed for exact circuit synthesis, the algorithm
may also be used to construct an $\epsilon$-approximate circuit.
}

\section{Conclusion}
\label{sec:conc}
We have presented a quantum synthesis algorithm designed to produce short circuits and scale well in practice. QFAST belongs in the class of bottom-up synthesis tools that use numerical optimization. The main contribution is a circuit encoding that allows replacing expansive search over large circuit spaces with a single step of numerical optimization in a topology aware manner. The evaluation on depth optimal circuits, as well as circuits generated by domain  generators (VQE, TFIM) indicates that while not optimal, QFAST can significantly reduce the depth of circuits used in practice by domain scientists. This reduction is beyond the capabilities of other existing synthesis tools or optimizing compilers.
This bodes well for the future adoption of synthesis for algorithm discovery or circuit optimization during the NISQ era and beyond.

\section*{Acknowledgments}

This work was supported by the DOE under contract DE-5AC02-05CH11231,
through the Office of Advanced Scientific Computing Research (ASCR)
Quantum Algorithms Team and Accelerated Research in Quantum Computing programs.

\bibliographystyle{abbrv}
\bibliography{bibliography,quantum,quant_chem}

\end{document}